\documentclass{ws-ijmpa}

\begin{document}

\markboth{M. V. Chizhov \& D. P. Kirilova} {Chiral tensor particles
in the early Universe}

%%%%%%%%%%%%%%%%%%%%% Publisher's Area please ignore %%%%%%%%%%%%%%%
%
%\catchline{}{}{}{}{}
%
%%%%%%%%%%%%%%%%%%%%%%%%%%%%%%%%%%%%%%%%%%%%%%%%%%%%%%%%%%%%%%%%%%%%

\title{Chiral tensor particles in the early Universe
%\footnote{For the title, try not to use more than
%3 lines. Typeset the title in 10 pt roman, uppercase and boldface.}
}

\author{M. V. Chizhov}

\address{Centre for Space Research and
Technologies,\\ Faculty of Physics, Sofia University, 1164 Sofia,
Bulgaria\\
%\footnote{State completely
%without abbreviations, the affiliation and mailing address,
%including country. Typeset in 8 pt italic.}\\
mih@phys.uni-sofia.bg}

\author{D. P. Kirilova}

\address{Institute of Astronomy,
Bulgarian Academy of Sciences,\\
BG-1784, Sofia, Bulgaria\\
dani@astro.bas.bg}

\maketitle

\begin{history}
\received{Day Month Year} \revised{Day Month Year}
\end{history}

\begin{abstract}
     The status of the chiral tensor particles in the extended electroweak
model, their experimental constraints, signatures and the
possibilities for their detection at the new colliders are reviewed.
The characteristic interactions of the chiral tensor particles in
the early Universe plasma and the corresponding period of their
cosmological influence is determined. The dynamical cosmological
effect, namely the speeding of the Friedmann expansion due to the
density increase caused by the introduction of the new particles, is
evaluated.

It is shown that the existence of the chiral tensor particles is
allowed from cosmological considerations and welcomed by the
particle physics phenomenology.

\keywords{chiral tensor particles; cosmology}
\end{abstract}

\ccode{PACS numbers: 95.30.Cq, 12.90.+b}

\section{Introduction}

In this work we will discuss  the presence of a new type of spin-1
particles in the primeval plasma and their effects on the Friedmann
expansion and on the processes in the early Universe.

%Many talks at this conference are dedicated to the very popular
%today modification of the left-hand side of the Einstein--Hilbert
%equation.
%Contrary to this we will leave it unmodified,
%the left-hand side,
%which leads to the conventional standard cosmology, while

Neither a supersymmetry, nor extra-dimensions have been observed or detected up
till now and they remain just a theoretical construction. Nevertheless, they
are considered as natural extensions of the standard model of elementary
particles and their search continues in most of the particle physics
experiments dedicated to physics beyond the standard model.

Physics beyond the standard model, discussed here, namely the
introduction of new chiral tensor (ChT) particles into the standard
electroweak model, is based on the experimental fact that such type
of particles exist in Nature at the QCD scale as the hadron
resonances.\cite{NJL}
%This extension proposes an extension of the
An extension of the standard model by ChT particles at the
electroweak scale is similar to the technicolor idea, which
considers hadron resonances like prototypes of the electroweak
bosons.

On the other hand, from theoretical point of view, the ChT particles
are inserted into the standard model of elementary particles to
complete the set of Yukawa interactions and realize all possible
irreducible representations of the Lorentz group. Such model was
first presented in Ref.~\refcite{MPLA} and was discussed in more
detail in the review in Ref.~\refcite{EChAYa}.

The new ChT particles, present in the early Universe, contribute to
the matter tensor in the right-hand side of the Einstein--Hilbert
equation, increasing the Universe density and changing the dynamical
evolution of the Universe. Besides they have direct interactions
with the particles present at the early stage of the Universe
evolution.

In the next section we briefly review the current status of the ChT
particles. In the third section we estimate  the characteristic
scale of their typical processes and their effect on the expansion
rate of the Universe and discuss the BBN constraint on their
effective coupling constant. The last section presents our
conclusions and vision about future exploration of the ChT
particles.

\section{Today's status of chiral tensor particles}

%\subsection{Main characteristics}

%The main characteristic feature of the new particles is  already
%denoted by their name.
According to the extended model\cite{MPLA,EChAYa} with ChT
particles, the latter are described by an {\em antisymmetric} tensor
fields of a rank two, like the stress tensor of the electromagnetic
field. An antisymmetric tensor field describes two fields with spin
one, which are represented by a polar and an axial vectors with
respect to spatial transformations. (Unlike the common association
of the word {\it tensor} with particles with spin two, which are
described by a second rank {\em symmetric} tensor, like the
graviton.)

{\it\bf Chirality}.
%is the characteristic feature  of the particles concerning
%their interactions. )
The new particles have a chiral charge and change the fermion
chirality in contrast to the gauge bosons with spin one, which have
minimal interactions with fermions and do not change their
chirality. They have an anomalous (Pauli) interaction with matter,
which provides a distinguishing signature for their detection.

%Why chiral tensor particles should be introduced in the theory

{\it\bf Introduction of the chiral tensor particles into the SM}.
The ChT particles are introduced as doublets $\left(T^+_{\mu\nu}
T^0_{\mu\nu}\right)$, like Higgs particles, due to the same as Higgs
chirality property. As a result of their richer interaction
possibilities new chiral anomalies appear. The latter are avoided by
introducing an additional doublet $\left(U^0_{\mu\nu}
U^-_{\mu\nu}\right)$ with an opposite chiral- and hypercharge. This
doubling of doublets concerns also the Higgs sector, thus, similar
to the SUSY case, it becomes $\left(H^+_1 H^0_1\right)$,
$\left(H^0_2 H^-_2\right)$.
%Thus, the complete set of the matter fields can be
%expressed by: $\left(H^+_1 H^0_1\right)$, $\left(H^0_2
%H^-_2\right)$, $\left(T^+_{\mu\nu} T^0_{\mu\nu}\right)$ and
%$\left(U^0_{\mu\nu} U^-_{\mu\nu}\right)$.
% where the standard model components are denoted in blue,
% while the introduced in the model ones are denoted in red.

Massless tensor particles have just longitudinal degrees of freedom.
In order for them to acquire mass a Higgs-like mechanism should be
applied. However, in this case the role of the Higgs field will be
played by a triplet and a singlet gauge vector particles or by four
$SU(2)_L$ singlets (depending on the chiralities of the ChT
particles), which supply them with the transverse physical degrees
of freedom. Thus, in addition to the discussed Higgs doublets an
extra triplet and five singlets, denoted further on by $C_{\mu}$ and
$P^i_{\mu}$ ($i=1,...,5$), correspondingly, should be introduced as
well for the two tensor doublets.

However, the doublets doubling may cause a flavor violation in the
neutral sector. This can be easily avoided if the doublets $H_1$ and
$T_{\mu\nu}$ interact only with down-type fermions, while the
doublets $H_2$ and $U_{\mu\nu}$ -- with up-type ones.\cite{2higgs}

{\it\bf Chiral tensor particles degrees of freedom}. The
antisymmetric tensor particles while massless, have only
longitudinal physical degrees of freedom opposite to the case of
gauge fields. Therefore, the presence of the two additional tensor
doublets, the triplet and singlets gauge vector particles and the
extra Higgs doublet leads to an increase of the total effective
number of the degrees of freedom while the additional particles are
relativistic by $g_{ChT}=g_T+g_U+g_C+g_P+g_H=4+4+6+10+4=28$, namely:
$g_{*}=g_{\rm SM}+g_{ChT}=106.75+28=134.75$.

 {\it\bf Chiral tensor particles masses}. As in the SM the
particle masses are induced
%at lower energies
through the Higgs mechanism. The presence of the vacuum expectation
values of the two different Higgs doublets leads to different masses
for the tensor particles interacting with up- and down-type
fermions, correspondingly $M_U\sim700$~GeV  and $M_T\sim 1$~TeV. The
spectrum of the ChT particle masses in the extended model is
considered more precisely in Ref.~\refcite{ICTP}.

{\it\bf Experimental signatures and constraints}. A. The presence of
the ChT particles do not contradict the present experimental data
and moreover can explain successfully a number of anomalies  in the
precise low energy physics, namely:
\begin{itemlist}
\item
{\it weak radiative pion decay anomaly:}
 The Dalitz plot of the weak
radiative pion decay cannot be interpreted in the framework of the
standard V-A interactions.\cite{Bolotov} The anomaly consists in the
fact that the predicted decay probability is higher than the
measured one. In the standard model the theoretically calculated
probability is defined by the sum of the squares of two matrix
elements. In the extended electroweak model with the ChT
particles\cite{discovery} there exists an additional matrix element
in the probability, which leads to a distinctive destructive
interference. Hence, the predicted probability matches the measured
one, provided the effective coupling constant of the tensor
interactions is  $G_T\sim 10^{-2}G_F$, where $G_F$ is the Fermi
coupling constant, i.e. the new interactions are centi-weak. A
cosmological constraint of the same order can be obtained from BBN
considerations
%bounds on right-handed neutrinos,
as discussed in section 3.
\item
{\it CVC anomaly in tau decays:} There exists  4.6 $\sigma$ difference between
the predicted from $e^+e^-$ annihilation and the measured branching ratio of
tau decay into two pions.\cite{Davier} This anomaly can be explained using the
lepton universality of the tensor interactions and the same strength $G_T\sim
10^{-2}G_F$, as found above.\cite{CVC}
\item
{\it muon g-2 anomaly:} There exists a 3.3 $\sigma$ discrepancy
between the measured and theoretically predicted value of the muon
anomalous magnetic moment.\cite{amu} The presence of the massive
neutral tensor particles can naturally explain this anomaly: Thanks
to the mixing of the neutral tensor particle with the photon, the
photon shows an anomalous interactions with the fermions, which
actually reflects the anomalous interaction of the tensor particles
with the fermions. Moreover, unlike other existing in literature
models explaining this anomaly, this mechanism predicts a systematic
shift in the fine structure constant value determined from the
electron anomalous magnetic moment.\cite{ae}
\end{itemlist}

B. Besides the successful explanation of the listed above anomalies,
the {\it tensor interactions do not contradict the precise low
energy experiments}  like: super-allowed beta decay, muon decay,
neutron decay, etc.\cite{lowenergy} The results of
DELPHI\cite{DELPHI}, TWIST\cite{TWIST} and $\mu_{P_T}$\cite{muT}
experiments do not exclude the presence of these particles with the
parameters discussed above.

The chiral tensor particles may be produced and detected at powerful
high energy colliders. For example, some feature indicative for the
ChT particles was already observed at the Tevatron \cite{Tevatron}.
The chiral tensor particles, if they exist, should be certainly
discovered at the ongoing Large Hadron Collider by CMS and ATLAS
experiments at CERN, in case they have the mentioned above masses
and coupling constants.\cite{LHC}

\section{Cosmological Effects of Chiral Tensor Particles}

Cosmological influence of antisymmetric tensor particles was first
discussed in Ref.~\refcite{Kirilova}. Here we use the currently
updated characteristics of the ChT particles to discuss their effect
on the early Universe. We will consider two types of cosmologically
important effects of ChT particles: first, the additional particles
influence on Universe dynamics,
%due to the  increase of the energy density. And
and second, ChT particles direct interactions with the other constituents of
the early Universe plasma.

{\it\bf ChT particles effect on the Universe expansion.} The
presence of two doublets of antisymmetric tensor particles and the
corresponding additional Higgs doublet, triplet and singlets of
gauge vector particles leads to a considerable increase of the
energy density of the Universe in comparison with the Standard
Cosmological Model case $\rho=\rho_{SCM}+\rho_{ChT}$. This results
in  speeding up the Friedmann expansion $H=\sqrt{\frac{8\pi}{3}\:
G_N\rho}$, where $H$ is the rate of the expansion, $G_N$ is the
gravitational constant. At the early stage, while the additional
particles are relativistic, their contribution can be expressed
through the effective degrees of freedom, namely:
$\rho_{ChT}=\frac{\pi^2}{30}\:g_{ChT}\: T^4$, where $T$ is the
photons temperature. Hence, for cosmic times $t$ later than ChT
particles creation time $t_c$ and before ChT particles became
non-relativistic or decay, i.e. $t_c<t<t_d$, the Friedmann expansion
is speeded up $H=\sqrt{8\pi^3 G_N g_*/90}\:T^2$, where $g_*$ is the
total number of the effective degrees of freedom $g_*=134.75$, as
discussed in the previous section. The temperature-time dependence
is also shifted, namely the cosmic time, corresponding to a given
temperature, slightly decreases compared to the standard
cosmological model case, since $t\sim1/(\sqrt{g_*}\:T^2)$ and
$g_*>g_{SM}$.

{\it\bf ChT particles interactions in the early Universe.} Through
their interactions with the fermions the tensor particles directly
influence the early Universe plasma. Analyzing the interactions of
the tensor particles, we have estimated the characteristic
temperatures and cosmic times for these particles, namely their
creation, scattering, annihilation and decay.

Tensor particle interactions become effective when the
characteristic rates of interactions $\Gamma_{int}\sim \sigma n$
become greater than the expansion rate $H(T)$. At energies greater
than the tensor particles masses, the cross sections have the
following behavior $\sigma\sim E^{-2}$. Hence at very high energies
tensor particles have been frozen, and with the cooling of the
Universe during its expansion, they unfreeze. The temperature of
unfreezing $T_{eff}$, i.e. the temperature corresponding to the
beginning of the epoch of particles effectiveness  due to a given
interaction $i\rightarrow f$ is defined from the relation
$\sigma_{if}(T)n(T)=H(T)$
%$T^2_{eff}=\sqrt{90/8\pi^3 G_N g_*}\:n\sigma_{if}$
and the corresponding cosmic time is estimated as $t_{eff}\approx
2.42/(\sqrt{g_*}\:T^2_{eff}$[MeV])~s.

The cross-section for the creation of pairs of longitudinal
\footnote{The interactions of the transverse components of the tensor fields are
similar to the interactions of the transversal gauge W and Z fields,
and therefore  we are not going to discuss them here.}
 tensor
particles from fermion-antifermion collisions is calculated to be:
\begin{equation}\label{cr}
    \sigma_c\approx\frac{\pi\alpha^2\ln(E/v)}{64\sin^4\theta_W E^2},
\end{equation}
where the fine-structure constant $\alpha(M_Z)\approx 1/127.9$, the weak-mixing
angle $\sin^2\theta_W\approx 0.23136$ and the Higgs vacuum expectation
$v\approx 250$~GeV. Thus the tensor particle creation becomes effective at
$T<T_c\approx 3.3\times 10^{16}$~GeV, which corresponds to cosmic times
$t>t_c\approx 1.9\times 10^{-40}$~s.

The fermions scattering on tensor particles has a cross-section
\begin{equation}\label{sc}
    \sigma_s\approx\frac{\pi\alpha^2}{192\sin^4\theta_W E^2}.
\end{equation}
It becomes effective at $T<T_s\approx 3.4\times 10^{14}$~GeV and
$t>t_s\approx 1.8\times 10^{-36}$~s.

Tensor particles annihilations proceed till $t_a\approx
2.42/(\sqrt{g_*}\:T_a^2{\rm [MeV]})~{\rm s}\approx 2.1\times
10^{-13}$~s, where  $T_a=2M_T$.

The decay width of the tensor particles is $\Gamma\approx \alpha
M_T/\sin^2\theta_W\approx 34$~GeV. The lifetime and the
corresponding cosmological temperature are $t_d=1.9\times
10^{-26}$~s and $T_d=3.3\times 10^9$~GeV. The decay time is much
earlier than the annihilation, hence the tensor particles disappear
from the cosmic plasma mainly by decaying. The period of their
effectiveness is the period from the time of their creation to the
moment of their decay, $t_c=1.9\times 10^{-40}~{\rm s}
<t<t_d=1.9\times 10^{-26}$~s,  i.e. during the very early stage of
the Universe evolution. The corresponding energy range spreads  from
$10^{16}$~GeV  down to $10^9$~GeV. So, the ChT particles decay
safely early so that their decay products do not disturb the  CMB.
On the other hand they are present at energies typical for
inflation, Universe reheating, lepto- and baryogenesis, et cetera.
The ChT particles eventual role in these processes is to be explored
in future studies. We would like only to mention  here that the
extended model with ChT particles proposes new source for
CP-violation due to the richer structure of particles and
interactions, and therefore, may present a natural mechanism for
leptogenesis and baryogenesis scenarios.

{\it\bf BBN constraint on the ChT interactions strength.} In the
discussed extended model with the ChT particles right-handed
neutrinos interact with the chiral tensor particles and in case the
neutrinos are light they can be produced through ChT particles
exchange. Then it is straightforward to obtain rough  cosmological
bound on the coupling constant of the ChT particles $G_T$ on the
basis of BBN considerations.\cite{Dolgov}

Using the BBN bound on the additional light neutrinos
$N_{eff}=g_R(T_{\nu_R}/T_{\nu_L})^4<1$ and assuming three light
right-handed neutrinos, it follows that
$3(T_{\nu_R}/T_{\nu_L})^4<1$, which puts a constraint on the
decoupling/freezing  of the right-handed neutrinos. The temperature
of freezing $T_f$ of $\nu_R$ may be determined using the BBN
constraint and entropy conservation relation $g T^3=const$, namely
$T_{\nu_R}/T_{\nu_L}=(\frac{43}{4}/g_*(T_f))^{1/3}<0.76$. Then the
rough constraint on the decoupling temperature is $T_f>140$ MeV.

On the other hand the decoupling temperature of a given species is
connected with its interactions coupling strength, hence
$(G_T/G_F)^2\sim(T_f/3~{\rm MeV})^{-3}$, where we assume $3$ MeV as
the decoupling temperature of the active neutrino species. Then the
constraint on the ChT coupling is $G_T<10^{-2}G_F$.

In case of two light right handed neutrinos, the corresponding
constraints on the decoupling temperature is $T_f>100$ MeV, which
slightly changes the constraint on the coupling. These two cases
provide BBN constraints  in agreement with the value of the $G_T$
provided from the experimental data discussed in the second section.
So BBN cosmological constraint points as well to the possibility of
a centi-weak tensor interactions.\footnote{Considerably looser BBN
bound will follow in case of only one light right-handed neutrino
species, as seen from the above considerations, namely $G_T\le G_F$.
 The eventual future detection of the ChT particles and determination of their
coupling constant may point to the number of the light right-handed
neutrino species.}

\section{Conclusion}

As has been discussed, the existence of chiral tensor  particles
does not contradict the experimental data of the precise low energy
experiments like: super-allowed beta decay, muon decay, neutron
decay, etc. Moreover, chiral tensor particles  help to explain some
anomalies in the precision low energy physics, namely, the weak
radiative pion decay anomaly, the CVC anomaly in tau decays, the
muon g-2 anomaly, etc. Besides, chiral tensor particle anomalous
interactions with matter provide a distinguishing signature for
their detection. The new particles may be produced and detected at
powerful high energy colliders. Particularly, if they exist, they
should be certainly discovered at the ongoing Large Hadron Collider
by CMS and ATLAS experiments at CERN, in case they have the
mentioned above masses and coupling constants. At present the search
of these particles is included in the ongoing experimental program
of the ATLAS Collaboration at LHC. Some feature indicative for
tensor particles was already observed at the
Tevatron.\cite{Tevatron}

Besides, as demonstrated in this work, these particles are allowed
by cosmology, as well. BBN cosmological considerations point to the
possibility of a centi-week tensor interactions. ChT particles cause
a slight increase of the Friedmann expansion and correspondingly
change the temperature-time dependence during the period of
effectiveness of their interactions with the other Universe
constituents: This period lasts from the time of their creation till
their decay, namely $t_c=1.9\times 10^{-40}~{\rm s}<t<t_d=1.8\times
10^{-26}$ s. The corresponding energy range is from $10^{16}$~GeV
(typical for the inflationary period) down to $10^9$~GeV, which
according to us is very promising for theoretical speculations
involving the chiral tensor particles concerning inflationary
models, reheating scenarios, baryogenesis and leptogenesis
scenarios, etc.

 Hopefully, in future more particle physics experiments will  pay attention
to the experimental exploration of the characteristics of the chiral
tensor particles and they will be included in the theoretical models
of the early Universe evolution as well.

\section*{Acknowledgments}

The work of M.V. Chizhov was financially supported by the
JINR--Bulgaria grant for 2009 year.

%\begin{thebibliography}{000} %for 3 digits
%\begin{thebibliography}{00}  %for 2 digits

\end{document}